\documentclass[conference]{IEEEtran}
\IEEEoverridecommandlockouts
\usepackage{cite}
\usepackage{amsmath,amssymb,amsfonts}
\usepackage{algorithmic}
\usepackage{graphicx}
\usepackage{textcomp}
\usepackage{xcolor}
\usepackage{url} 
\usepackage{tabularx}
\usepackage{makecell}
\usepackage{caption}
\def\BibTeX{{\rm B\kern-.05em{\sc i\kern-.025em b}\kern-.08em
    T\kern-.1667em\lower.7ex\hbox{E}\kern-.125emX}}

\newboolean{showcomments}
\setboolean{showcomments}{true} 
\ifthenelse{\boolean{showcomments}}
  {\newcommand{\nb}[2]{
    \fcolorbox{gray}{yellow}{\bfseries\sffamily\scriptsize#1}
    {$\blacktriangleright$#2$\blacktriangleleft$}
   }
   
  }
  {\newcommand{\nb}[2]{}
   
  }

\usepackage{cleveref}

\usepackage{flushend}

\begin{document}

\title{The Feasibility of a Smart Contract ``Kill Switch''\\
}

\author{\IEEEauthorblockN{Oshani Seneviratne}
\IEEEauthorblockA{\textit{Computer Science Department}, \textit{Renssealer Polytechnic Institute}\\
\textit{Troy, New York, USA} \\
senevo@rpi.edu}
}

\maketitle

\thispagestyle{plain}
\pagestyle{plain}

\begin{abstract}

The advent of blockchain technology and its adoption across various sectors have raised critical discussions about the need for regulatory mechanisms to ensure consumer protection, maintain financial stability, and address privacy concerns without compromising the foundational principles of decentralization and immutability inherent in blockchain platforms. We examine the existing mechanisms for smart contract termination across several major blockchain platforms, including Ethereum, BNB Smart Chain, Cardano, Solana, Hyperledger Fabric, Corda, IOTA, Apotos, and Sui. We assess the compatibility of these mechanisms with the requirements of the EU Data Act, focusing on aspects such as consumer protection, error correction, and regulatory compliance. Our analysis reveals a diverse landscape of approaches, from immutable smart contracts with built-in termination conditions to upgradable smart contracts that allow for post-deployment modifications. We discuss the challenges associated with implementing the so-called smart contract ``kill switches,'' such as the balance between enabling regulatory compliance and preserving the decentralized ethos, the technical feasibility of such mechanisms, and the implications for security and trust in the ecosystem.

\end{abstract}


\section{Introduction}

Blockchain technology offers unprecedented opportunities for innovation, efficiency, and trust. At the heart of this transformation are smart contracts — autonomous, self-executing agreements embedded in code, which have the potential to redefine interactions within various sectors, from finance and healthcare to supply chain management and beyond. However, integrating these technologies into the fabric of societal systems raises complex regulatory, ethical, and operational challenges. Among these is the necessity to reconcile the inherently decentralized and immutable nature of blockchains with the evolving landscape of global regulations to ensure consumer protection, privacy, and the stability of financial systems.

Chen et al.~\cite{chen2020defining} noted that many smart contracts deployed on blockchains have security, availability, performance, maintainability, and reusability problems. According to Perez et al.~\cite{perez2021smart}, there are as many as 23,327 vulnerable contracts on the Ethereum platform alone, putting millions of dollars worth of cryptocurrencies owned by unsuspecting users in jeopardy.
Therefore, it is unsurprising that smart contract ``kill switch'' requirements from regulatory bodies have emerged. 

The European Union's Data Act, specifically Article 30~\cite{EUDataAct2023}, proposes the concept of a ``kill switch" to empower authorities and possibly participants within blockchain ecosystems to intervene directly in the operation of smart contracts — a concept that, at first glance, seems at odds with the principles of decentralization and immutability that define blockchain technology. 
However, although the EU Data Act presents an important vision for enhancing smart contracts, we must critically assess its practicality and desirability.
Implementing robust smart contract termination or interruption may have several logistical challenges, as smart contracts are fixed in content and operation at the time of deployment and essentially follow the ``Code is Law'' ethos~\cite{de2018blockchain}.
In this paper, we explore various pathways for developing smart contract standards for ``kill switches'' that can accommodate regulatory expectations without compromising the unique advantages of blockchains. 

The remainder of this paper is organized as follows: \Cref{sec:background} provides background information on the regulatory framework, arguments for and against smart contract regulation, potential applications of smart contract ``kill switches'' across different domains, and a review of related work. \Cref{sec:existing-solutions} examines existing blockchain solutions and their suitability for implementing smart contract ``kill switches.'' \Cref{sec:discussion} discusses the effects on current ecosystems. Finally, \Cref{sec:conclusion} concludes the paper and suggests future research directions.

\section{Background}
\label{sec:background}

The smart contract ``kill switch'' concept has garnered significant attention in academic literature and industry discussions. This attention stems from the increasing realization of the potential risks and challenges associated with deploying immutable and autonomous smart contracts, especially in critical financial, legal, and social applications. 
Article 30 of the Data Act~\cite{EUDataAct2023} focuses specifically on requirements concerning smart contracts used in a data spaces context. 
The proposal sets out four requirements for smart contracts to make data available: (1) robustness, (2) safe termination and interruption, (3) data archiving and continuity, and (4) access control.
According to the Act, platform providers and individuals deploying smart contracts for data-sharing purposes must ensure that the smart contract is robust against errors or malicious attacks, protected via rigorous access control mechanisms, and can be terminated or interrupted. The smart contact data, logic, and code can be archived to facilitate auditing if terminated. 

\subsection{Problem Addressed with Smart Contract ``Kill Switches''}

Smart contracts represent a significant advancement in blockchain technologies and are still part of an emerging field.
Once the smart contracts are executed, they cannot be unilaterally intercepted or modified, even if the underlying contract is deemed void or unenforceable. 
As summarized in \Cref{table:comparative_works}, there is a general lack of flexibility, dependence on oracles, vulnerability to bugs and architectural changes (such as the infamous Ethereum DAO hack and the aftermath~\cite{dhillon2017dao}), immutability and privacy concerns, and enforcement issues.
There are also complexities with smart contracts when they diverge from their legal intentions, emphasizing the challenge of unwinding or terminating them when needed~\cite{meyer2020stopping}. 
It is crucial to establish a clear distinction between a smart contract as a technical tool and the legal contract it represents. This distinction highlights the difficulties in aligning the programmed actions of smart contracts with the mutable and often subjective nature of legal interpretations and expectations.

In the broader context of the current work, which discusses ``kill switches'' as a regulatory and safety mechanism in smart contracts, it is imperative for smart contracts to incorporate mechanisms that allow for legal intervention and adjustments. This is crucial for legal compliance without compromising the decentralized and automated nature of blockchains.

Implementing such mechanisms involves weighing various benefits and potential drawbacks.
There are several pros and cons of terminating a smart contract based on an external trigger, as outlined in \Cref{table:kill_switch_pros_cons}.

\begin{table}[ht]
\caption{Pros and Cons of ``Kill Switches''}
\centering
\begin{tabularx}{\linewidth}{|>{\hsize=0.2\hsize}X|>{\hsize=0.4\hsize}X|>{\hsize=0.4\hsize}X|}
\hline
\textbf{Aspect} & \textbf{Pros} & \textbf{Cons} \\
\hline
Security & Enhances protection against vulnerabilities and bugs. & Potential target for malicious actors if not securely managed. \\
\hline
Compliance & Facilitates compliance with regulations like the EU Data Act. & May conflict with the principle of immutability in blockchains. \\
\hline
Governance & Can be designed to involve community consensus. & Might introduce elements of central control. \\
\hline
User Trust & Increases confidence in safety mechanisms. & Users may fear misuse or overreach. \\
\hline
\end{tabularx}
\label{table:kill_switch_pros_cons}
\end{table}

\subsection{Potential Applications}
\label{sec:applications}

\begin{figure}[t]
    \centering
    \includegraphics[width=\columnwidth]{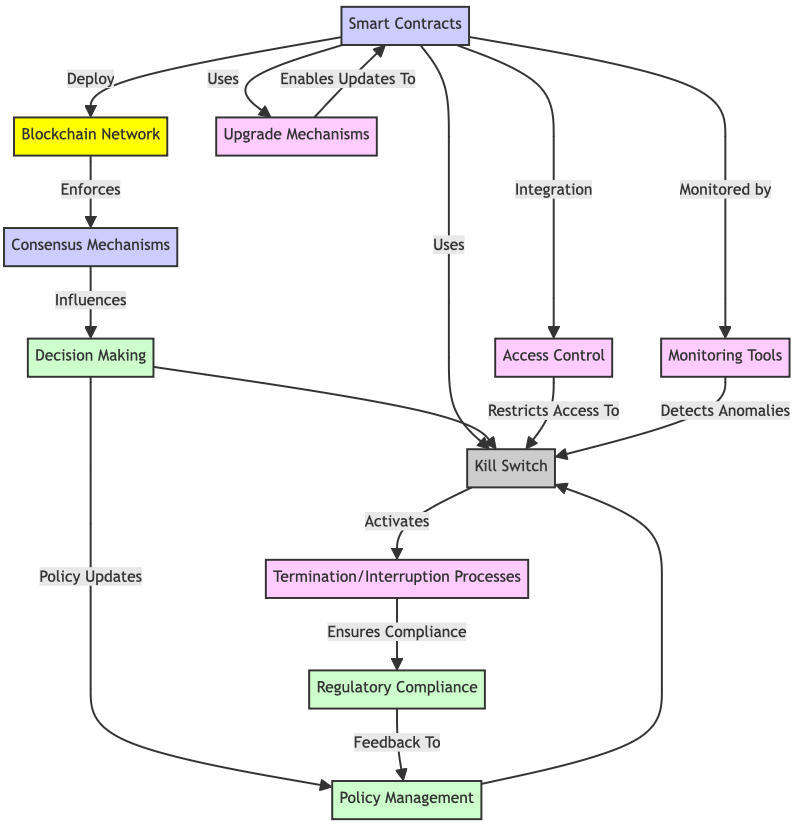}
    \caption{Relational Graph for a Smart Contract ``Kill Switch'' Implementation}
    \label{fig:workflow}
\end{figure}

Smart contract ``kill switches'' have a wide array of potential applications across various industries, offering a valuable tool for enhancing security, compliance, and operational flexibility. 
\Cref{fig:workflow} outlines the various components involved in managing the lifecycle and compliance of smart contracts in an idealistic environment, highlighting the interconnected roles of governance, technology, and monitoring that might be necessary for implementing such a ``kill switch'' mandated by regulation.

Domain-specific applications may emerge around the utility of ``kill switches'' in various industries beyond the legislative interest. \Cref{table:kill_switch_applications} outlines current and potential applications with some support for pausing and terminating the application.

\begin{table}[ht]
\caption{Potential Applications of ``Kill Switches''}
\label{table:kill_switch_applications}
\centering
\begin{tabularx}{\linewidth}{|>{\hsize=0.18\hsize}X|>{\hsize=0.38\hsize}X|>{\hsize=0.44\hsize}X|}
\hline
\textbf{Domain} & \textbf{Application} & \textbf{Purpose} \\
\hline
Finance & Decentralized Finance (DeFi) platforms involving stablecoins and other financial instruments. & Freezes transactions or adjusts parameters during market crashes, suspicious activities, or security breaches~\cite{li2024stablecoin}. \\
\hline
Healthcare & Smart contracts managing sensitive patient data or automated drug delivery systems such as the BlockIoT system~\cite{shukla2021blockiot,shukla2021blockiot-retel}. & Protects privacy by terminating contracts in case of data breaches in compliance with regulations like HIPAA, possibly utilizing standards-based ontological concepts for any unexpected situations warranting a pause~\cite{li2019leveraging}. \\
\hline
Supply Chain Management & Contracts for tracking payloads with robotic agents managed with smart contracts~\cite{grey2020swarm}. & Halts operations in response to detected anomalies in the operating environment~\cite{mallikarachchi2022managing}. \\
\hline
\end{tabularx}
\captionsetup{justification=centering, singlelinecheck=false}
\end{table}


    



\subsection{Related Work} 
\label{sec:related-work}

\Cref{table:comparative_works} outlines the key contributions of several related that address smart contract termination solutions.
In contrast to these works, we provide a comparative analysis of smart contract termination mechanisms across several major blockchain platforms in \Cref{sec:existing-solutions}. We specifically address the implementation challenges, governance models, and impact on decentralization, which these previous studies have not covered comprehensively.

\begin{table}[ht]
\centering
\caption{Related Work Comparison}
\begin{tabularx}{\linewidth}{|>{\hsize=0.15\hsize}X|>{\hsize=0.5\hsize}X|>{\hsize=0.35\hsize}X|}
\hline
\textbf{Study} & \textbf{Key Contributions} & \textbf{Gaps} \\
\hline
Casolari et al.~\cite{casolari2023improve} & Examine the role of smart contracts in the architecture of the EU's Data Act, identifying key challenges and proposing recommendations to address those issues. & Specific mechanisms for smart contract termination in various blockchains are lacking. \\
\hline
Olivieri and Pasetto \cite{olivieri2023towards} & Analysis of EU Data Act requirements for smart contracts, focusing on interoperability, robustness, and safe termination. & Specific mechanisms for smart contract termination in various blockchains are lacking. \\
\hline
Le et al. \cite{le2018proving} & Method for proving conditional termination of smart contracts utilizing the F* programming language.  Before execution, the system checks those conditions against the current state and inputs to decide whether the contract can safely run without leading to non-termination scenarios. & Limitations in automatically inferring termination proofs for complex programs, necessitating manual intervention for complex cases. \\
\hline
Genet et al. \cite{genet2020termination} & Formal and mechanized proof of termination based on measures of EVM call stacks for intrinsic system-wide safeguards (gas and call stack limits). & Comparative analysis of termination mechanisms across different blockchains is lacking. \\
\hline
Liu et al. \cite{liu2019strengthening} & Strengthening Hyperledger Fabric Chaincode smart contracts to handle unexpected situations, which is unlocked through a novel voting algorithm. & Only applicable to private-permissioned blockchains, and the sandbox environment for voting may not be practical. \\
\hline
Zhu et al. \cite{zhu2019proposal}) & Recovering any ``lost'' crypto tokens after a voting round empirically shown to be resilient in the face of any Sybil attacks and adversarial collusion. & Generalizability of the proposed method to other smart contract termination scenarios is questionable, and the sandbox environment for voting may not be practical. \\
\hline
Mohsin et al. \cite{mohsin2019ontology} & Utilizing community-accepted off-chain ontologies as a guiding framework for action in case of anomalies or errors in deployed contracts. & Ontology as a decision-support mechanism requires strong governance and trust guarantees. \\
\hline
Marino et al. \cite{marino2016setting} & Legal frameworks for altering and undoing smart contracts. & Solution through pure legal means may impact decentralization and user trust. \\
\hline
\end{tabularx}
\label{table:comparative_works}
\end{table}

\section{Existing Solutions} 
\label{sec:existing-solutions}

We outline approaches for smart contract termination already available in several prominent blockchains and how they could support the EU Data Act mandate for smart contract ``kill switch'' in \Cref{table:kill_switch_approaches_eu_data_act}. 
We compiled this table upon the examination of some of the prominent blockchains that support smart contracts along several dimensions, including the following:

\begin{enumerate}
    
    \item \textbf{Strategy}: The methods and strategies used by the blockchain platform to implement ``kill switches'' in smart contracts, which could include built-in functions, design patterns, or other relevant features.

    \item \textbf{Strengths}: The inherent strengths of the platform for smart contract termination.

    \item \textbf{Weaknesses}: The inherent weaknesses of the platform for smart contract termination.



    \item \textbf{Governance}: (Abbreviated to \textit{Gov.} in \Cref{table:kill_switch_approaches_eu_data_act}) Captures whether any governance mechanisms or protocols within the blockchain allow network participants to intervene or make decisions regarding the termination or pausing of smart contracts. 
    

    \item \textbf{Regulation Support}: Discusses the potential or existing support for compliance with regulatory frameworks, specifically the European Union Data Act. 
\end{enumerate}

\begin{table*}[t]
\caption{Comparison of Smart Contract ``Kill Switch'' Approaches in Various Blockchain Implementations}
\begin{tabularx}{\linewidth}{|>{\hsize=0.1\hsize}X|>{\hsize=0.25\hsize}X|>{\hsize=0.2\hsize}X|>{\hsize=0.2\hsize}X|>{\hsize=0.05\hsize}X|>{\hsize=0.15\hsize}X|}
\hline
\textbf{Blockchain} & \textbf{Strategies} & \textbf{Strengths} & \textbf{Weaknesses} & \textbf{Gov.} & \textbf{Regulation Support} \\
\hline
Ethereum~\cite{buterin2013ethereum} \& BNB Smart Chain~\cite{BNBChainWhitepaper} & Self-destruct function included in the Solidity language; Pause and emergency stop design patterns; Upgradeable contracts. & Provides built-in functions for contract termination; Compatible with the widespread tools and infrastructure. & No external mechanism; Potential security risks; Possible removal of the self-destruct function raises concerns about long-term viability. & No & Yes, through custom implementations using the Solidity features. \\
\hline
Cardano~\cite{CardanoWhitepaper} & Design-specific conditions within smart contracts built into Plutus; Stateful smart contracts; Seamless interaction with off-chain code. & Uses a robust functional programming language (Haskell) for Plutus; Strong on-chain governance mechanisms. & No external mechanism; Complex implementation and limited adoption compared to Ethereum. & Yes & Yes, through design-specific conditions. \\
\hline
Solana~\cite{yakovenko2018solana} & Upgradable programs; State management. & High throughput and low latency with upgradable programs. & No external mechanism; Immaturity of the ecosystem and less community support for governance models. & No & Yes, through upgradable programs. \\
\hline
Hyperledger Fabric~\cite{androulaki2018hyperledger} & Chaincode lifecycle management; Endorsement policies; Private data collection; Administrative control. & Permissioned blockchain with strong lifecycle management and administrative controls. & Centralized nature might not align with decentralization principles. & Yes & Yes, through administrative control and governance mechanisms. \\
\hline
Corda~\cite{brown2016corda} & Built-in contract upgrade; Explicit termination conditions; Administrative control. & Focus on privacy and business transactions with upgradable contracts. & Limited use cases outside of enterprise applications. & Yes & Yes, through explicit contract conditions. \\
\hline
IOTA~\cite{saa2023iota} & State management built into the ISCP; Ability to respond to external inputs or triggers that could include termination signals. & Scalable with no transaction fees suitable for IoT. & Still evolving with ongoing updates to smart contract capabilities. & Yes & Yes, through decentralized control mechanisms. \\
\hline
Aptos~\cite{AptosWhitepaper2022} \& Sui~\cite{SuiWhitepaper2023} & Move language flexibility for contract updates; Expressive smart contract implementations tracking and managing assets. & Strong type system for formal verification and security; Supports more complex governance and transaction models. & Newer ecosystems with less mature tooling and support. & Yes & Yes, through explicit contract conditions. \\
\hline
\end{tabularx}
\label{table:kill_switch_approaches_eu_data_act}
\end{table*}

\subsection{Ethereum}

In Ethereum~\cite{buterin2013ethereum}, smart contract termination and interruption are primarily handled through the built-in functionalities of the smart contracts themselves. Ethereum does not provide an external ``kill switch'' or mechanism for forcibly terminating or interrupting smart contracts from outside the contract's code. Instead, the implementation of such features is left to the developers who write the smart contracts, typically managed through the following mechanisms:

\begin{itemize}
    
    \item \textbf{Self-Destruct Function:} This function (originally called SUICIDE) allows a contract to be terminated, removing its code and storage from the blockchain~\cite{chen2021smart}. When a contract is self-destructed, it sends the remaining Ether stored to a designated address and removes the code from the blockchain, making it inoperable. However, the contract's code and past transactions are immutable and still part of the blockchain history.
    This function is typically used to remove contracts that are no longer needed or recover funds in an emergency. It must be explicitly included in the smart contract code. It can only be triggered by a function call within the contract, often restricted to the contract owner or other authorized entities.
    There is a recent proposal on removing this function~\cite{ButerinSelfdestruct}, as it is the only opcode that breaks important invariants, which causes an unbounded number of state objects to be altered in a single block. Therefore, the long-term availability of this functionality is uncertain.

    \item \textbf{Pause and Emergency Stop Patterns:} For interruption rather than complete termination, EVM-based smart contracts can be designed with pause or emergency stop functionalities~\cite{EthereumDevPausingContracts}. These patterns allow certain contract functions to be turned off temporarily without removing the contract from the blockchain, which can be useful when a bug is discovered and the contract needs to be paused to prevent further damage while a fix is being developed.
    The pause pattern typically involves setting a boolean variable that controls the execution of sensitive functions. By changing this variable's state, the contract's critical operations can be enabled or disabled. The emergency stop pattern is more comprehensive, allowing for a phased approach to pausing and resuming contract functionalities, often with different levels of access control and conditions for triggering and reversing the pause state~\cite{SolidityEmergencyStop}.

    \item \textbf{Upgradeable Contracts:} Another approach to managing smart contract behavior over time, including termination and interruption, is through upgradeable contracts~\cite{salehi2022not}. This design pattern involves deploying a proxy contract that delegates calls to an implementation contract containing the logic. If the implementation needs to be changed, updated, or fixed, a new implementation contract can be deployed, and the proxy contract is updated to delegate calls to the new contract. This approach allows bugs to be fixed and functionalities to be updated without terminating the contract. However, it may introduce complexity and potential security considerations.

\end{itemize}



Other popular public permissionless blockchains, such as BNB Smart Chain (BSC)~\cite{BNBChainWhitepaper}, formerly known as Binance Smart Chain, is a blockchain platform that runs parallel with Binance Chain. It offers smart contract functionality and compatibility with Ethereum's existing infrastructure, such as the Ethereum Virtual Machine (EVM). This compatibility allows it to support Ethereum tools and DApps, making it a popular choice for developers looking to leverage the scalability and performance benefits of BSC while maintaining access to Ethereum's rich ecosystem.
Handling smart contract termination and interruption in BNB Smart Chain is similar to Ethereum, primarily because of its EVM compatibility. 

\subsection{Cardano}

Cardano~\cite{CardanoWhitepaper} is a blockchain platform that employs a layered architecture. It separates the settlement layer, which handles transactions, from the computational layer, where smart contracts run. Cardano uses a unique proof-of-stake consensus algorithm called Ouroboros and supports smart contracts through its native programming language, Plutus~\cite{CardanoPlutus}.
Plutus is designed to enable the creation, execution, and management of smart contracts on the Cardano blockchain. Plutus contracts are written in Haskell, a functional programming language known for its high fault tolerance and security features. The use of Haskell influences how smart contracts, including their termination and interruption, are handled in Cardano.

\begin{itemize}
    \item \textbf{Termination by Design:} In Cardano, the termination or interruption of a smart contract is primarily a matter of the contract's design. Because Plutus allows for creating highly deterministic and secure contracts, developers can incorporate specific conditions under which a contract may terminate or pause its operations. These conditions are encoded directly into the contract's logic and can be triggered by predefined events or states.

    \item \textbf{Stateful Smart Contracts:} Cardano's smart contracts can manage the state through the blockchain ledger, but how the state is handled is distinct from other platforms. Termination or modification of a contract could involve creating transactions that update or end the contract's state according to the logic defined in the contract itself, which ensures that the contract's behavior remains predictable and tamper-proof.

    \item \textbf{Off-chain Code:} Cardano also supports off-chain code execution through its application framework, which allows for complex interactions with on-chain smart contracts. Interruptions or terminations initiated by off-chain components can be designed to interact with the on-chain contracts, offering another layer of control for managing contract lifecycles. This off-chain logic can facilitate scenarios where user interaction or external data triggers the pause or stop conditions in the smart contract.

    \item \textbf{Governance and Updates:} Cardano's governance model can play a role in contracts that require the ability to evolve or might need to incorporate mechanisms for interruption or termination post-deployment. Through on-chain governance mechanisms, stakeholders can propose and vote on updates or changes to smart contracts, assuming the contract is designed to be upgradable, and the governance model supports such actions. This approach allows the community or stakeholders to have a say in the contract's lifecycle management.

\end{itemize}

\subsection{Solana}

Solana~\cite{yakovenko2018solana} is a high-performance blockchain platform designed to support scalable, decentralized applications and cryptocurrencies. It uses a unique consensus mechanism called Proof of History (PoH), combined with the underlying Proof of Stake (PoS) consensus, to achieve high throughput and low latency. Unlike Ethereum and other blockchains, where smart contract termination and interruption mechanisms are more explicitly discussed and implemented, Solana's approach to smart contract management, including termination and interruption, is somewhat different due to its architecture and programming model.
In Solana, smart contracts are referred to as ``programs." These programs are written in Rust or C, compiled to Berkeley Packet Filter (BPF) bytecode, and deployed to the Solana blockchain. Once deployed, a program can be interacted with by sending transactions from Solana accounts, but it is immutable, which means there is no built-in ``kill switch'' or termination mechanism for a Solana program once it is live on the network, but termination-like behavior can be achieved through the following:

\begin{itemize}
    \item \textbf{Upgradable Programs:} Solana has a mechanism for program upgradability through the use of a ``Program Upgradeable Loader"~\cite{solanabpfloader}, which allows developers to deploy a new version of a program to replace the old one. The process involves deploying the new program version as a separate entity and then ``switching" the program authority to point to the new program. This method does not terminate the old program but effectively redirects interactions to the new, upgraded program version.

    \item \textbf{State Management:} Termination or interruption of a program's operation in the traditional sense may not directly apply to Solana's model. However, programs can manage their state through accounts that hold data. By modifying the state held in these accounts, a program can implement mechanisms to halt or modify its operations based on specific conditions, essentially allowing for a form of ``interruption" of its functions.

\end{itemize}

\subsection{Hyperledger Fabric}

Hyperledger Fabric~\cite{androulaki2018hyperledger} uses a permissioned blockchain platform designed primarily for enterprise use. Hyperledger Fabric refers to smart contracts as ``chaincode." The smart contract termination and interruption approach in HyperLedger Fabric is characterized by its lifecycle management features, endorsement policies, and the control mechanisms provided by its permissioned network structure. Collectively, these features offer a structured and governed way to manage chaincode operations, including their update, interruption, and termination, in line with the needs and policies of the enterprise blockchain network.

\begin{itemize}
    
    \item \textbf{Chaincode Lifecycle Management:} Hyperledger Fabric introduces sophisticated lifecycle management for chaincodes~\cite{chaincodelifeccycle}, allowing organizations to agree on chaincode parameters before deployment to the network. This lifecycle management process enables more granular control over the deployment, upgrade, and management of chaincode, including their termination and interruption. Hyperledger Fabric also allows upgrading the chaincode contract to a new version by deploying the new contract on the network and performing an upgrade transaction. The upgrade can introduce new logic, fix issues, or modify the chaincode's behavior. This process is controlled and requires consensus from the participating organizations, ensuring that changes are agreed upon before implementation.

    \item \textbf{Chaincode Endorsement Policies:} Hyperledger Fabric employs endorsement policies~\cite{chaincodeendorsement} define the rules under which a transaction is considered valid. These rules could include those that might terminate or interrupt chaincode operations. Chaincode can require that transactions be endorsed by a specific number of peers from certain organizations within the network, offering a high level of control and security over chaincode execution, including any operations that could stop or alter the chaincode's function.

    \item \textbf{Private Data Collections:} Hyperledger Fabric supports private data collections~\cite{chaincodeprivatedata}, which allow a subset of the network to transact privately, maintaining confidentiality. If such a chaincode contract is updated or removed, the data governed by the policies of the private data collection remains, ensuring that sensitive information is handled according to the requirements, even if the chaincode's operation is interrupted or terminated.

    \item \textbf{Administrative Operations:} Due to the permissioned nature of Hyperledger Fabric, network administrators have more control over the chaincode contracts, including their deployment, operation, and termination. Therefore, if necessary, chaincode contracts can be administratively stopped or removed by parties with the appropriate permissions, according to the governance model of the specific Hyperledger Fabric network.

\end{itemize}

\subsection{Corda}

Corda's architecture and operational model offer unique mechanisms for managing the lifecycle of `Corda Contracts"~\cite{brown2016corda}.
Corda's design emphasizes privacy and finality in transactions, influencing how contract termination and interruptions are perceived and managed. Transactions in Corda are only shared with parties directly involved or who need to validate them. Once a transaction is finalized, it is considered immutable and authoritative, aligning with business needs for certainty and finality in agreements.

Corda handles smart contract termination and interruption through its contract upgradeability features, contract constraints governing state evolution, and explicitly modeling termination logic within contract code. Its architecture supports the management of the contract lifecycle in a way that aligns with the platform's focus on direct, private, and final transactions among business entities.

\begin{itemize}
    \item \textbf{Upgradability:} Corda provides a built-in mechanism for contract upgradability, which allows the network participants to evolve their contracts over time as business needs change or in response to discovering issues with the original contract. Upgrading a contract in Corda involves transitioning the states governed by an old version of the contract to a new version under the agreement of all relevant parties.

    \item \textbf{Contract Constraints:} Corda uses a concept called ``contract constraints" to govern which contract codes can constrain the evolution of ledger states. These constraints ensure that once states are created under a specific contract, future transactions that consume and evolve these states are validated by the same contract code or an agreed-upon upgraded version, providing a form of governance over contract changes.

    \item \textbf{Explicit Termination and State Evolution:} Contracts can be designed to include termination logic or conditions within their clauses. Since contracts in Corda govern the transition of states, a contract can explicitly define conditions under which a state is considered final or can no longer be evolved, effectively terminating the contract's applicability to that state. Additionally, business processes can be modeled to include explicit termination transactions that move states to a final, consumed status, where they cannot be used in future transactions.

    \item \textbf{Flow Framework:} Corda's Flow Framework~\cite{cordaflow}, which facilitates the automation of transactions between nodes, can be used to manage the execution of contract termination or state evolution logic. Through flows, participants can coordinate complex processes, including those involving contract or state termination, under the rules defined by their Corda contracts.

    \item \textbf{Administrative Intervention:} In a permissioned network like Corda, network operators have administrative control over the network, including the ability to intervene in the operation of contracts and nodes in accordance with the network's governance policies. This process includes managing membership and potentially coordinating contract upgrades or the resolution of disputes related to contract execution.

\end{itemize}

\subsection{IOTA}

IOTA~\cite{saa2023iota} is a blockchain designed primarily for the Internet of Things (IoT) environment, focusing on scalability, speed, and the elimination of transaction fees. Unlike public permissionless networks like Ethereum or permissioned networks like Hyperledger Fabric, IOTA utilizes a unique data structure called the Tangle~\cite{popov2018tangle}, which is a form of Directed Acyclic Graph (DAG) that facilitates different operational characteristics and advantages, particularly in terms of scalability and transaction fees. 
IOTA introduced smart contracts as part of its ecosystem to provide more complex and conditional transaction capabilities through the IOTA Smart Contracts Protocol (ISCP).
ISCP operates on the second layer on top of the IOTA Tangle, providing the flexibility needed for complex computations and smart contracts that wouldn't be feasible directly on the Tangle due to its structure aimed at handling transactions efficiently. This adaptability ensures that ISCP can handle a wide range of smart contract scenarios, providing reassurance to developers and users alike.
In ISCP, smart contracts run on their separate chains, known as ``chain accounts," which are independent but anchored to the main IOTA Tangle. This design allows for greater scalability, as each smart contract can operate on its own chain without overwhelming the main network.
Smart contracts in IOTA can define their validators (known as committee nodes), who are responsible for executing the contract and reaching a consensus on its state. This design allows contract creators to tailor the security and consensus mechanisms to their needs, balancing decentralization, security, and efficiency.
With ISCP, developers can program smart contracts in Rust, a language known for its safety and performance. This choice underlines the focus on creating secure and efficient smart contracts capable of supporting various applications, from DeFi to IoT.

It is worth noting that the IOTA project has undergone significant updates and expansions to its technology stack, aiming to address various challenges and expand its use cases beyond the IoT. These updates include enhancements to smart contract functionalities, interoperability features, and scalability solutions, which may influence how smart contract termination and interruption are handled in future iterations.

\subsection{Aptos and Sui}

More recent entries into the field of blockchains that utilize DAGs, such as Aptos~\cite{AptosWhitepaper2022} and Sui~\cite{SuiWhitepaper2023}, are making notable advancements by adopting the Move programming language~\cite{blackshear2019move} for their smart contract functionality. The move language, designed with safety and security as its core principles, caters directly to the needs of financial applications and services by enabling a precise definition of custom resource types. These resources are linear types that cannot be copied or implicitly discarded, ensuring assets are tracked and managed securely throughout their lifecycle. Move's ability to define resource types aligns well with the transactional requirements of these DAG-based blockchains, allowing for more expressive and flexible smart contract implementations compared to traditional scripting languages. This design choice not only reduces the likelihood of bugs that lead to significant vulnerabilities (such as reentrancy attacks) but also opens up possibilities for implementing more complex governance and transaction models that can adapt over time while maintaining rigorous security and integrity standards.

\section{Discussion}
\label{sec:discussion}

As illustrated in \Cref{sec:existing-solutions}, platforms like Ethereum and BNB Smart Chain utilize smart contract-level features such as self-destruct functions and pause patterns, while others, like Hyperledger Fabric and Corda, rely more heavily on governance and administrative controls to manage contract lifecycle and termination.
Platforms like Cardano and Corda stand out for their emphasis on design-specific conditions and explicit contract terms built into their corresponding smart contract languages. This approach, which is a blend of technical foresight and governance oversight, provides a robust framework for smart contract development.
Therefore, it is evident that integrating smart contract``kill switch" mechanisms is influenced by both program-defined measures and governance-based solutions. 

The type of blockchain (public, private, or consortium) and the consensus mechanisms also play crucial roles in determining the feasibility and implementation style of  smart contract ``kill switches.'' Public blockchains require broader consensus for changes, which can complicate the rapid deployment of robust smart contract termination mechanisms. However, private and consortium blockchains can implement these features more straightforwardly due to their centralized governance structures.

Integrating ``kill switches'' into smart contracts has broad implications for the blockchain ecosystem. These mechanisms, designed to intervene in unforeseen circumstances or malfunctions, raise debates concerning decentralization, asset management, and security. 
Some of these implications are as follows:

\begin{itemize}
    \item \textbf{Concerns Regarding True Decentralization}: One of the foundational principles of blockchain technology is decentralization — the idea that any single entity does not control operations and governance. Introducing ``kill switches'' into smart contracts presents a paradox; while they can provide necessary safety nets for users, they also introduce a vector for centralized control. Critics argue that this undermines the very essence of decentralization~\cite{cointelegraph2023}. However, it's essential to recognize that many blockchains already incorporate mechanisms for updates and upgrades (as noted in \Cref{sec:existing-solutions}), some of which require centralized decision-making or a coordinated consensus among stakeholders. 

    \item \textbf{Loss of Assets}: Activating a smart contract ``kill switch'' could potentially lead to scenarios where users lose access to their assets temporarily or permanently. This risk is particularly acute in financial applications where smart contracts govern the custody and transfer of significant value. Therefore, any ``kill switch'' implementation must include safeguards to prevent unintentional or unjustified wiping out of value. Such safeguards could involve mechanisms for restoring operations and assets post-intervention, transparent and fair criteria for activation, and perhaps insurance mechanisms to cover losses in the worst-case scenarios. 

    \item \textbf{Security Issues}: Implementing smart contract ``kill switches'' introduces specific security considerations, particularly regarding the key management or permissions required to activate or deactivate the switch. If not managed securely, these could become targets for malicious actors looking to disrupt operations and exploit the assets secured by the smart contract. It's suggested that separate keys or permissions be used for the activation (pausing) and deactivation (unpausing) processes to minimize risks. Furthermore, these keys should be rotated or changed once used to prevent reuse attacks. 

\end{itemize}

\section{Conclusion} 
\label{sec:conclusion}

We explored the feasibility and implications of implementing a smart contract ``kill switch" mechanism within the framework of blockchains in light of the European Union's Data Act legislation~\cite{EUDataAct2023}. 
Our findings contribute to the ongoing debate on regulating blockchain technology, providing insights into how current blockchain platforms can adapt to meet legislative requirements without stifling innovation and be accessible and understandable by non-technical users. 
The discourse around smart contract ``kill switches'' is multifaceted, reflecting a cross-section of academic, legislative, and industry perspectives. 

The challenge lies in designing ``kill switch'' mechanisms that align with the ethos of decentralization as much as possible, perhaps through decentralized governance models or community consensus mechanisms.
We believe a hybrid model where decentralized platforms can interact with regulatory frameworks without compromising their decentralized nature is necessary for a smart contract ``kill switch'' to take effect successfully.
Adopting ``kill switches'' in smart contracts within the blockchain ecosystem demands careful consideration of their impacts on decentralization, asset security, and the broader trust in blockchain technologies. By addressing these concerns thoughtfully, it's possible to design systems that retain the benefits of decentralization while providing mechanisms to protect users and the integrity of the network. This process involves a delicate balance between control and freedom, requiring ongoing dialogue and innovation within the community to navigate these complex issues effectively.

Future studies could explore the design, implementation, and effectiveness of decentralized governance models specifically tailored to manage smart contracts ``kill switch'' mechanisms. 
Investigating automated mechanisms within smart contracts that dynamically adjust to changing regulatory requirements without manual intervention could be a significant area for exploration, particularly for already deployed smart contracts.
It may be necessary to include protocol updates through hard forks or the governance models of various blockchain projects that allow for changes to be made to operational parameters. 
A deeper analysis of how ``kill switch'' mechanisms affect the security, trust, and overall perception of blockchain networks among users before and after implementing ``kill switches'' and security vulnerability assessments related to their deployment would provide insights into the long-term feasibility of smart contract termination solutions.
Additionally, with the increasing diversity of blockchain platforms, there is a need to focus on developing cross-chain solutions and interoperability standards that facilitate regulatory compliance across different blockchains. 
Future research is likely to dig deeper into these discussions, proposing frameworks, models, and real-world trials that balance the autonomy of smart contracts with the safety, security, and compliance requirements of the broader ecosystem.

\medskip

\noindent\textbf{Acknowledgments:} The author acknowledges the support from NSF IUCRC CRAFT center research grant (CRAFT Grant \#22018) for this research. The opinions expressed in this publication do not necessarily represent the views of NSF IUCRC CRAFT.
The author would also like to thank Sabrina Kirrane for her valuable initial input that contributed to the development of this paper.

\bibliographystyle{ieeetr}
\bibliography{references}

\end{document}